\documentstyle[12pt]{article}

\newwrite\ffile\global\newcount\figno \global\figno=1

\def\writedef#1{}

\input epsf
\def\figin{\epsfcheck\figin}\def\figins{\epsfcheck\figins}
\def\epsfcheck{\ifx\epsfbox\UnDeFiNeD
\message{(NO epsf.tex, FIGURES WILL BE IGNORED)}
\gdef\figin##1{\vskip2in}\gdef\figins##1{\hskip.5in}
\else\message{(FIGURES WILL BE INCLUDED)}%
\gdef\figin##1{##1}\gdef\figins##1{##1}\fi}
\def\figinsert{}
\def\ifig#1#2#3{\xdef#1{fig.~\the\figno}
\writedef{#1\leftbracket fig.\noexpand~\the\figno}%
\figinsert\figin{\centerline{#3}}\medskip\centerline{\vbox{\baselineskip12pt
\advance\hsize by -1truein\center\footnotesize{  Fig.~\the\figno.} #2}}
\bigskip\endinsert\global\advance\figno by1}
\def\endinsert{}

\begin{document}
\begin{titlepage}
\flushright{IP/BBSR/2002-06}

\vspace{1in}

\begin{center}
\Large
{\bf Logarithmic Corrections to Black Hole Entropy and AdS/CFT
Correspondence}
\vspace{1in}

\normalsize

\large{ Sudipta Mukherji and Shesansu Sekhar  Pal }\footnote{e-mail: mukherji, 
shesansu@iopb.res.in}

\normalsize
\vspace{.7in}

{\em Institute of Physics \\
Bhubaneswar - 751005, India }

\end{center}

\vspace{1in}

\def    \beq    {\begin{equation}}

\baselineskip=24pt
\begin{abstract}
We calculate the correction to the Bekenstein-Hawking entropy formula 
for five dimensional AdS-Schwarzschild black holes due to thermodynamic 
fluctuations. The result is then
compared with the boundary gauge theory entropy corrections via 
AdS/CFT correspondence. We then further generalise our analysis for the
rotating black hole in five dimensional AdS space.

\end{abstract}

\end{titlepage}
In recent years, new insights of space-time physics have emerged
particularly through realising the idea of holography in terms 
of AdS/CFT correspondence \cite{Maldacena}~-\cite{agmoo}. According to 
this
correspondence, type IIB
string theory in a five dimensional AdS space times a five-sphere 
is dual to the large $N$ limit of ${\cal N}$ = 4 supersymmetric Yang-Mills
theory in four dimensions. More interestingly, in \cite{witten}, 
it was further argued that the gauge theory in question, at high
temperature, is dual to AdS-Schwarzschild black hole at the same
temperature. This indeed leads to a new interpretation of 
Hawking-Page phase transition in terms of  transition between the
confining and de-confining phase of Yang-Mills gauge theory.
It further follows from \cite{witten} that the thermodynamic
variables of CFT can be read off from the boundary of AdS-Schwarzschild
black hole. In particular, at high temperature, CFT calculation gives the
correct entropy of the black hole up to a constant numerical
multiplicative factor.

Recently, there are several works in the literature \cite{kmone}~-
\cite{bs} 
suggesting that for
a large class of black holes (AdS-Schwarzchild being one), 
the Bekenstein-Hawking area-entropy law receives additive logarithmic 
corrections due to thermal fluctuations. Typically, the corrected
formula has the form
\begin{equation}
{\cal S} = S_0 - c~ {\rm log} S_0,
\label{correc}
\end{equation}
where $S_0$ is the standard Bekenstein-Hawking term and $c$ is a
number. The purpose of this
note is to analyse this correction of the entropy of AdS-Schwarzschild 
black hole in the light of AdS/CFT correspondence. In particular, in this
note, we  calculate the correction to the entropy of  CFT due
to thermal fluctuations and then compare it with the entropy correction of
the black hole (\ref{correc}) following AdS/CFT dictionary.
 
The logarithmic correction to the entropy due to thermal 
fluctuation comes out to be of the form given is (\ref{correc})
with $c = 1/2$. We show that this matches with what is expected from 
AdS/CFT correspondence. However, this number differs from what was found 
in existing literature, see for example \cite{bdm}
\footnote{ We thank Parthasarathi Majumdar for several fruitful e-mail 
correspondence with us in this regard.}. 

In \cite{bdm}, the correction to the Bekenstein-Hawking entropy 
of AdS-Schwarzschild black hole has been calculated within canonical
ensemble.  We first briefly review the calculation in grand canonical
setup
\footnote{However, later in the calculation, we will 
set the chemical potential to zero, making this equivalent to canonical 
ensamble}. 


The partition function in the grand canonical ensemble is given by
\begin{equation}
Z(\alpha,\beta)=\int^{\infty}_0\int^{\infty}_0 \rho(n,E)e^{(\alpha
  n-\beta E)} dn dE,
\label{gcpart}
\end{equation}  
where $\alpha=\beta\mu$, $\mu$ is the chemical potential and
$\beta=\frac{1}{T}$ is the inverse temperature. Here we have set 
$k_B=1$ so that the temperature is having the dimension of energy. 
The density of states, $\rho(n,E)$, can be obtained from the
above equation by inverse Laplace transformation\cite{BohrMottelson}, 
as
\begin{eqnarray}
\rho(n,E)&=&(\frac{1}{2\pi i})^2\int^{c+i\infty}_{c-i\infty}
\int^{c+i\infty}_{c-i\infty} Z(\alpha,
\beta) e^{(-\alpha n+\beta E)}d\alpha d\beta,\nonumber \\& &
=(\frac{1}{2\pi i})^2\int^{c+i\infty}_{c-i\infty}\int^{c+i\infty}_{c-i\infty}
e^{S(n,E,\alpha,\beta)}d\alpha d\beta,
\label{invlap}
\end{eqnarray}
where the function $S(n,E,\alpha,\beta)$ is defined  through
\begin{equation}
S(n,E,\alpha,\beta)={\rm ln} Z(\alpha,\beta)-\alpha n+\beta E.
\end{equation} 
The integral in (\ref{invlap}) will be performed by the saddle-point
approximation and the
main
contribution to this integral will come around the equilibrium point
$(\alpha_0, \beta_0)$, where the integrand is stationary, and the
stationary conditions are determined by $\frac{\partial
  S(n,E,\alpha,\beta)}{\partial\alpha}|_{(\alpha_0,\beta_0)}=0$, and 
$\frac{\partial
  S(n,E,\alpha,\beta)}{\partial\beta}|_{(\alpha_0,\beta_0)}=0$, which
amounts to the following two equations
\begin{eqnarray}
& &\frac{\partial {\rm ln}
Z}{\partial\alpha}|_{(\alpha_0,\beta_0)}-n=0,\nonumber \\& &
\frac{\partial {\rm ln} Z}{\partial\beta}|_{(\alpha_0,\beta_0)}+E=0.
\end{eqnarray}
Expanding the exponent in the integrand of 
(\ref{invlap}) to second order around the
point $(\alpha_0,\beta_0)$, we get a Gaussian integral.
Finally, after performing the integral we obtain
\begin{equation}
\rho(n,E)=\frac{e^{S(\alpha_0,\beta_0)}}{2\pi \sqrt{M}},
\end{equation}
where $M$ is the determinant of the following matrix:
\begin{equation}
M=\left( \begin{array}{cc} 
                          \frac{\partial^2 {\rm ln}Z}{\partial\alpha^2}|_{(\alpha_0,\beta_0)} &
                          \frac{\partial^2 {\rm ln}Z}{\partial\alpha\partial\beta}|_{(\alpha_0,\beta_0)} \\
                          \frac{\partial^2 {\rm ln}Z}{\partial\beta\partial\alpha}|_{(\alpha_0,\beta_0)} &
\frac{\partial^2 {\rm ln}Z}{\partial\beta^2}|_{(\alpha_0,\beta_0)}  
                          \end{array} \right) \
\end{equation}
Evaluating the determinant (with the constraint that the
values of the off diagonal elements of the matrix $M$ are small in
comparison
to the diagonal elements) we  get the following from of the density
of states:
\begin{equation}
\rho(n,E)=\frac{e^{S(\alpha_0,\beta_0)}}{2\pi 
\sqrt{\frac{\partial^2 {\rm ln}Z}{\partial\alpha^2}|_{(\alpha_0,\beta_0)}\times
\frac{\partial^2 {\rm ln}Z}{\partial\beta^2}|_{(\alpha_0,\beta_0)}}}.
\end{equation} 
Hence, the entropy is
\begin{equation}
{\cal S} \equiv
{\rm ln}({\cal {E}}\rho)=S(\alpha_0,\beta_0)+{\rm
ln}\frac{{\cal {E}}}{\sqrt{\frac{\partial^2
      {\rm ln}Z}{\partial\beta^2}|_{(\alpha_0,\beta_0)}}}+
\rm{higher~order~terms}.  
\label{en}
\end{equation} 
As can be seen that $\rho$ in our calculation has a dimension of inverse 
temperature. To get entropy as a log of dimensionless quantity, we have 
multiplied the density of state with ${\cal{E}}$ which has the dimension 
of energy. This scale ${\cal{E}}$ is set by the particular system in 
question.
Since $\frac{\partial^2 {\rm ln}Z}{\partial\beta^2}|_{(\alpha_0,\beta_0)}
=C_vT^2$, where $C_v$ is the specific heat at constant volume, the
correction to the entropy becomes:
\begin{equation}
{\cal S}=S(\beta_0)+{\rm ln }\frac{{\cal {E}}} 
{\sqrt{C_vT^2}}+\ldots
\label{entro1}
\end{equation}
In (\ref{en}) and (\ref{entro1}), we have set the chemical potential to 
zero. Furthemore, the scale $\cal{E}$ has been set to be the 
temperature $T$ of the system.  This is because we have temperature as 
the only 
available scale in canonical ensamble. Hence, the above equation becomes 

\begin{equation}
\label{entro}
{\cal S}=S(\beta_0)-\frac{1}{2}{\rm ln }{C_v}+\ldots 
\end{equation}
\\
 
Having obtained the general form of correction to entropy, we would like
to apply it for black holes. Note that, for (\ref{entro}) to make sense,
$C_v$ has to be positive. In the context of black hole, it means that the
variation of mass with respect to temperature is positive. Five
dimensional AdS-Schwarzschild black hole does indeed satisfy this property 
in certain choice of parameters. The metric of such black hole is given by
\begin{equation}
ds^2=-(1-\frac{16\pi MG_5}{3\Omega_3
  r^2}+\frac{r^2}{l^2})dt^2+(1-\frac{16\pi MG_5}{3\Omega_3 r^2}+
\frac{r^2}{l^2})^{-1}dr^2+r^2d\Omega^2_3,
\end{equation}
where $d\Omega^2_3$ is the metric on unit sphere $S^3$ and $\Omega_3$
is the area of that unit sphere. $G_5$, $M$ and $\Lambda = - 6/l^2$ are 
five dimensional Newton's constant, mass and the cosmological constant.
The horizon of the black hole is at 
\begin{equation}
r^2=r^2_+=\frac{l^2}{2}{\Big(-1+\sqrt{1+\frac{48\pi
MG_5}{3\Omega_3l^2}} \Big)}.
\end{equation}
The entropy,  temperature, and the specific heat are given by
\begin{eqnarray}
& &S_{BH}=\frac{\Omega_3r^3_+}{4G_5},\nonumber \\& & 
T=\frac{4r^2_++2l^2}{4\pi
  l^2r_+}=\frac{1}{\pi}(\frac{\Omega_3}{4G_5})^{\frac{1}{3}}\times
(1+\frac{2r^2_+}{l^2}) S^{-\frac{1}{3}}_{BH},\nonumber \\& &
C=\frac{dM}{dT_H}=\frac{\partial M}{\partial r_+}\frac{\partial
  r_+}{\partial T_H}=3\frac{2r^2_++l^2}{2r^2_+-l^2} S_{BH}.
\label{para}
\end{eqnarray}
If
\begin{equation}
r^2_+ >\frac{l^2}{2},
\end{equation}
the specific heat becomes positive and evaluating 
 the mass, temperature and specific heat of black hole, in this case, becomes:
\begin{eqnarray}
& &M\sim S^{\frac{4}{3}}_{BH},\nonumber \\& &
T_H=\frac{1}{\pi}(\frac{\Omega_3}{4G_5})^{\frac{1}{3}}\times
  \frac{2r^2_+}{l^2}S^{-\frac{1}{3}}_{BH}\sim
S^{\frac{1}{3}}_{BH},\nonumber \\& &
C\sim3S_{BH}.
\end{eqnarray}
On substituting the above values of mass, temperature and specific heat 
into the equation (\ref{entro}), we get the corrected entropy as
\begin{equation}
{\cal S}_{BH}=S_{BH}-\frac{1}{2}{\rm ln} S_{BH}+\ldots
\label{blackentro}
\end{equation}  

According to AdS/CFT correspondence, on the boundary of AdS, we have 
conformally invariant ${\cal N}=4$ U(N) SYMs theory. The particle content of
this theory are: $N^2$ gauge fields, $6N^2$ massless scalars and $4N^2$ Weyl
fermions, that is, we have $8N^2$ bosonic and $8N^2$ fermionic
degrees of
freedom. The free energy and hence the entropy can be calculated in
the free field limit $g^2_{YM}N\rightarrow 0$.  The free energy
of a gas of bosons and fermions in a box of volume $V^{(n)}$,  the
superscript denotes the  number of
spatial dimensions, calculated in the grand canonical ensemble is
\begin{equation}
F=-T ~{\rm ln} Z=T\sum_{s_i, p} s_i{\rm ln}(1-s_ie^{-\beta p}),
\end{equation} 
where $s_i$ is +1 for bosons and -1 for fermions. Now taking the
volume of the box to be very large implies $\sum_p=V^{(n)}\int\frac{ d^n
  p}{(2\pi)^n}$. The free energy, then,  becomes \cite{gkt}
\begin{eqnarray}
F&= &TV^{(n)}\int\frac{d^n p}{(2\pi)^n}\sum_{s_i} {\rm ln}(1-e^{(-\beta
p+{\rm ln}
  s_i)})\nonumber \\& &
=-T^{n+1}V^{(n)}/2^n\frac{1}{(\pi)^{n/2}}
\frac{\Gamma(n+1)}{\Gamma(n/2+1)}(2-2^{-n})\zeta(n+1).
\end{eqnarray}
For ${\cal N}=4$, U(N) SYM theory in four dimension,  the free energy is
\begin{equation}
F=-\frac{\pi^2N^2T^4V^{(3)}}{6}.
\end{equation} 
Evaluating the specific heat from the free energy using the
definitions as:


\begin{eqnarray}
& &E=F-T\Big(\frac{dF}{dT}\Big)_V,\nonumber \\
& &C_{CFT}=\Big(\frac{d E}{dT}\Big)_V,
\end{eqnarray}

and substituting in equation (\ref{entro}),  we get:
\begin{equation}
{\cal S}_{CFT}=S_{CFT}-\frac{1}{2}{\rm ln}(S_{CFT})+~{\rm constant},
\label{gaugeentro}
\end{equation} 
where 
\begin{equation}
S_{CFT}=\frac{2\pi^2N^2T^3V^{(3)}}{3}.
\label{gag}
\end{equation}
Here, $T = T_{CFT}$ is clearly associated to the temperature of the
gauge theory that resides on the boundary.


We would now like to compare the corrected entropy formula for the 
bulk (\ref{blackentro}) and for the boundary CFT (\ref{gaugeentro})
following AdS/CFT prescription. 
In order to do so, let us concentrate at an asymptotic distance $r\equiv
L >> r_+$ of the  AdS-Schwarzschild black
hole  where lives the gauge theory. In this region, the geometry of the
above space-time becomes
\begin{equation}
ds^2 \sim L^2[-\frac{dt^2}{l^2}+d\Omega^2_3].
\end{equation} 
Due to red-shift, the temperature at this boundary therefore is
\begin{equation}
T_{CFT}=\frac{T_{BH}}{\sqrt{-g_{00}}}=\frac{l}{L}T_{BH},
\label{tcft}
\end{equation}
where $T_{BH}$ is the Hawking temperature associated with the black hole.
Substituting (\ref{tcft}) in (\ref{para}) and using 
$N^2$ from the AdS/CFT dictionary \cite{Maldacena}, 
\begin{equation}
N^2=\frac{\pi l^3}{2G_5},
\end{equation}
we get the standard \footnote{ appearance of this extra $4/3$ factor in
the relation between entropy formulas is discussed in \cite{gkt}.}
\begin{equation}
S_{CFT}=\frac{2\pi^2r^3_+}{3G_5}=\frac{4}{3} S_{BH}.
\end{equation}  
This, in turn, implies that (\ref{gaugeentro}) can be written as
\begin{equation}
{\cal S}_{CFT}=\frac{4}{3}S_{BH}-\frac{1}{2} {\rm
  ln}(\frac{4}{3}S_{BH})+~{\rm constant}
\label{mod_entro}
\end{equation}
\\
Let us now consider the case of rotating black hole in AdS space.
The line element is \cite{hhr},
\begin{eqnarray}
ds^2 & = & - {\Delta \over \rho^2} \left( dt - {a \sin^2 \theta \over \Xi
} d\phi 
\right) ^{\! \! 2} + {\Delta_\theta \sin^2 \theta \over \rho^2} \left( a
\, dt - {\left( r^2 +
a^2
\right) \over \Xi} d\phi \right) ^{\! \! 2} \nonumber \\ & & + {\rho^2
\over \Delta}
dr^2 +
{\rho^2 \over \Delta_\theta} \, d \theta^2  + r^2 \cos^2 \theta \, d \psi
^2 \; ,
\label{ds2}
\end{eqnarray}
where $0 \leq \phi , \psi \leq 2 \pi$ and $0 \leq \theta \leq \pi / 2$,
and
\begin{eqnarray}
\Delta & = & \left( r^2 + a^2 \right) \left( 1 + {r^2 \over l^2} \right) -
2MG_5
\nonumber \\
\Delta_\theta & = & 1 - {a^2\over l^2} \cos^2 \theta \nonumber \\ \rho^2 &
= &
r^2
+
a^2
\cos^2 \theta \nonumber \\ \Xi & = & 1 - {a^2 \over l^2} \; .
\end{eqnarray}
The horizon is at 
\begin{equation}
r_+^2 = {l^2\over 2} \left ( - (1 + {a^2\over l^2}) 
+ {\sqrt{ ( 1 + {a^2\over l^2})^2 + 4 {(2 M G_5 -a^2)\over l^2}}}\right).
\end{equation}
The entropy and temperature associated with the black hole is given by
\begin{equation}
S_{BH} = {\pi^2 (r_+^2 + a^2) r_+\over {2 G_5 \Xi}}, \, \,
T_{BH} = {r_+ ( 1 + {a^2\over l^2} + 2 {r_+^2\over l^2})\over {2 \pi
(r_+^2 + a^2 )}}.
\end{equation}
The angular velocity at the horizon ${\Omega}$ and angular momentum
$J$ are given by
\begin{equation}
\Omega_H = {a \Xi \over{ r_+^2 + a^2}}, \, \, J = {\pi M a\over{2 \Xi^2}}.
\end{equation}
In the region, $a \rightarrow 0$, the specific heat at constant angular
momentum can easily be calculated and is given by
\begin{equation}
C = {3 \pi^2 r_+^5 \{9 J^2 (1 + r_+^2) + 2 (2r_+^2 + l^2) \tilde M^2\}
\over { 4 G_5 r_+^2 (2r_+^2 -l^2)\tilde M^2 + 9G_5 J^2 (r_+^2 + 3 l^2)}},
\label{rotspec}
\end{equation}
where $\tilde M$ is the mass of the rotating black hole above the 
AdS background and is given by
\begin{equation}
\tilde M = {3\pi\over {4 \Xi}} M.
\end{equation}
It follows clearly from (\ref{rotspec}), that for large enough $\tilde M$, 
there is a domain in which $C$ is positive. Now, using 
(\ref{rotspec}) and (\ref{entro}),  along with the bulk-boundary 
correspondence for rotating black hole in AdS space \cite{bp}, it 
is easy to check that in high temperature limit, the
relation (\ref{mod_entro}) re-emerges.\\


To conclude, we have analysed the correction to entropy due to 
thermal fluctuations. We then applied our result to AdS black holes
and compared it with what is expected from AdS/CFT correspondence. 
We found that the corrected entropy formula is of the form (\ref{correc}) 
with $c = 1/2$.
%

\bigskip

\bigskip

\noindent {\bf Acknowledgements: } It is a pleasure to thank Somendra 
Bhattacharjee, Dileep
Jatkar and Parthasarathi Majumdar for useful conversations.

\end{document}